\documentclass{aa}

\usepackage{graphicx}
\usepackage{txfonts}
\usepackage[rflt]{floatflt}
\usepackage{natbib}        % to use BibTex for A&A
\bibpunct{(}{)}{;}{a}{}{,} % to follow the A&A style

\newcommand{\base}{\ensuremath{\mathrm{B{}}}}

\newcommand{\vis}{\ensuremath{\mu{}}}

\newcommand{\tf}{\ensuremath{\mu_0{}}}

\newcommand{\dUD}{\ensuremath{\small{\varnothing{}}}}
\newcommand{\airy}{\ensuremath{\mathcal{A}}}

\newcommand{\alphaHya}{$\alpha$~Hya}
\newcommand{\aHya}{a~Hya}     
\newcommand{\nVel}{N~Vel}     
\newcommand{\deltaVir}{$\delta$~Vir}
\newcommand{\chiVir}{$\chi$~Vir}
\newcommand{\iotaHya}{$\iota$~Hya} 
\newcommand{\BKVir}{BK~Vir}
\newcommand{\epsCrt}{$\epsilon$~Crt}

\begin{document}
%%%%%%%%%%%%%%%%%%%%%%%%%%%%%%%%%%%%%%%%%%%%%%%%%%%%%%%%%%%%%%%%%%%
%% DOCUMENT TITLE
\title{Post-processing the VLTI fringe-tracking data:\\ First measurements of stars\thanks{Based on observations collected at the European Southern Observatory, Paranal, Chile. Public data can be downloaded at: \texttt{archive.eso.org/wdb/wdb/eso/pkg\_cont/query?pkg\_id=96433}}}
\titlerunning{Measuring stellar diameters with the VLTI fringe-tracking data.}
%% DOCUMENT AUTHORS
\author{
  J.-B.~Le~Bouquin\inst{1}
  \and R.~Abuter\inst{2}
  \and P.~Haguenauer\inst{1}
  \and B.~Bauvir\inst{2}
  \and D.~Popovic\inst{2}
  \and E.~Pozna\inst{2}
}

%% DOCUMENT AUTHORS INSTITUTES
\institute{
  European Southern Observatory, Casilla 19001, Santiago 19, Chile
  \and
  European Southern Observatory, Karl-Schwarzschild-Str.\ 2, 85748 Garching, Germany
}
%% DOCUMENT OFF PRINTS
\offprints{J.B.~Le~Bouquin\\
  \email{jlebouqu@eso.org}}
%% DOCUMENT DATE
\date{Received 16 July 2008; Accepted 10 October 2008}
%% ABSTRACT
% optional context
\abstract
{At the Very Large Telescope Interferometer, the purpose of the fringe-tracker FINITO is to stabilize the optical path differences between the beams, allowing longer integration times on the scientific instruments AMBER and MIDI.}
% aims
{Our goal is to demonstrate the potential of FINITO for providing $H$-band interferometric visibilities, simultaneously and in addition to its normal fringe-tracking role.}
% methods
{We use data obtained during the commissioning of the Reflective Memory Network Recorder at the Paranal observatory. This device has permitted the first recording of all relevant real-time data needed for a proper data-reduction.}
% results
{We show that post-processing the FINITO data allows valuable scientific visibilities to be measured. Over the several hours of our engineering experiment, the intrinsic transfer function is stable at the level of $\pm2$\%. Such stability would lead to robust measurements of science stars even without the observation of a calibration star within a short period of time. We briefly discuss the current limitations and the potential improvements.}
% optional conclusions
{}

%% KEYWORDS
\keywords{Techniques: high angular resolution - Techniques: interferometric - Instrumentation: high angular resolution - Instrumentation: interferometers}%
%% MAKE TITLES
\maketitle

%%%%%%%%%%%%%%%%%%%%%%%%%%%%%%%%%%%%%%%%%%%%%%%%%%%%%%%%%%%%%%%%%%%
\section{Introduction}

The purpose of the FINITO facility at the Very Large Telescope Interferometer \citep[VLTI, see][]{Scholler-2006jul} is to compensate for the random optical delay (OPD) introduced by atmospheric turbulence. If not corrected, such perturbation would make the interferometric fringes jitter on the detector, preventing the use of long integration time on scientific instruments. FINITO was offered to the astronomical community in September 2007, and since then several AMBER \citep{Petrov-2007mar} programs have already benefited from the enhanced data quality produced by fringe-tracking \citep{LeBouquin-2008apr,Wittkowski-2008feb}.

Nevertheless, FINITO can potentially tackle another important issue at VLTI: measuring robust, absolutely calibrated, interferometric visibilities in the near-IR. Indeed, AMBER has been designed to be a highly rewarding spectro-interferometer \citep{Petrov-2007mar,Weigelt-2007mar,Domiciano-2007mar,Meilland-2007mar}, but this has required technical trade-offs. It typically reads out more slowly than an atmospheric coherence time, and so is critically sensitive to fringe stabilization, and hence to seeing and mechanical vibrations. This issue is exacerbated by VLTI service-mode only providing a few calibration stars per night, which is insufficient for properly sampling the time variations of the instrumental response (called transfer function). All together, this makes the \emph{absolute} calibration of AMBER fringe contrast difficult \citep{Millour-2007may}.

We have recently carried out the necessary developments to make FINITO able to provide visibility measurements. No hardware modifications were needed as FINITO already interferometrically combined the light of 2 or 3 telescopes to stabilize the OPD. The main issue was the high frame rate of the system running generally at more than 1kHz. This was overcome by the commissioning of the so-called VLTI \emph{Reflective Memory Network Recorder} (RMNrec), which listens to the communications between several systems of VLTI and is able to store the real-time data into proper FITS files.

In this paper, we investigate the possibility of post-processing these real-time data to measure interferometric visibilities and to properly calibrate them. In Sect.~\ref{sec:observationsAndDataReduction} we summarize the FINITO and RMNrec instrumental setup, the test observations made, and the simple data reduction used for this demonstration. In Sect.~\ref{sec:resultsAndDiscussion}, as a classical test, we compare the measured angular diameter of some well-known stars with previously published values. We discuss the results, the accuracy obtained, and the perspectives for reaching higher precision in the future. The paper ends with brief conclusions and perspectives.

%%%%%%%%%%%%%%%%%%%%%%%%%%%%%%%%%%%%%%%%%%%%%%%%%%%%%%%%%%%%%%%%%%%
\section{Observations and data reduction}
\label{sec:observationsAndDataReduction}

\subsection{RMNRec and FINITO instrumental setup}

\begin{figure} % finito.fig
  \hspace{0.05cm} \includegraphics[scale=.65]{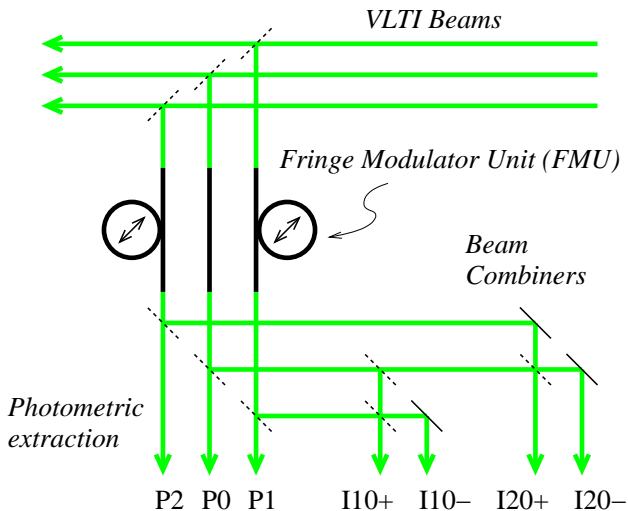}
  \centering \caption{\label{fig:finito} Sketch of the Finito optical bench showing the fibered Fringe Modulator Units (FMU) and the two co-axial beams combiners. Black solid lines are for the mirrors, while dashed lines are for semi-transparent devices.  \{$I_{10}^+$,$I_{10}^-$,$I_{20}^+$,$I_{20}^-$\} are the interferometric outputs and \{$P_1$,$P_0$,$P_2$\} are the photometric outputs. The real-time flux at those seven outputs are recorded in real time in a camera not displayed here.}
\end{figure}

For our study, we use the commissioning data from the RMNrec machine at Paranal. By looking at the Reflective Memory Network at high frequency, this computer constructs FITS files with the real-time data of several VLTI guiding and tracking facilities: error vector of the centroids as measured on the IRIS guiding camera; real-time fringes, phase and flux as measured on FINITO; and fringe-tracking loop status (search, idle, lock) and OPD offsets as handled by the OPD-Controller software (OPDC). Our long-term goal is to benefit for all information in the data processing of VLTI instruments. For the purpose of this paper, we only focus on the FINITO and OPDC data because they are of scientific interest in themselves, since they contain the interferometric visibility of the observed target.

A detailed description of the FINITO facility can be found in \citet{Gai-2004oct}. Recent updates can be found in \citet{LeBouquin-2008spie_b}. Briefly, FINITO interferometrically combines two pairs of telescopes from a telescope triangle in the $H$-band  (see Fig.~\ref{fig:finito}). The optical path length of two of the three beams are modulated internally using a piezo-driven expansion ring changing the length of the fibers, called a Fringe Modulator Unit (FMU). We generally use a triangular modulation of 5 $H$-band fringes in amplitude peak-to-peak and a sampling of 4 readouts per fringes. Each modulation ramp is hereafter called a scan. Photometric signals are extracted by Glan-Taylor polarizers, while the other polarization component is fed to a three-way, pairwise, co-axial beam combiner. The central beam (labelled 0) is split into equal intensity fractions and then combined with the modulated beams (respectively 1 and 2). This produces two pairs of complementary interferometric outputs ($\pi$ phase difference), for the combination 0-1 and 0-2, respectively. The 3 photometric signals ($P_1$,$P_0$,$P_2$) and the 4 interferometric signals ($I_{10}^+$,$I_{10}^-$,$I_{20}^+$,$I_{20}^-$) are routed to a Rockwell PICNIC camera, which is read at a frequency between 500Hz and 2kHz. These output signals are processed in real time to measure the fringe phases, used by the OPDC to reject fringe motion for frequencies below 20Hz.

\subsection{Observations}

Commissioning RMNrec data were taken during the nights of 2008-01-28, 2008-01-29, and 2008-01-30 with the interferometric baseline A0-D0 alone. FINITO was therefore tracking on a single channel (beams 0-2, known to provide better performances). Sequences of 3 files of $30$s were recorded at each telescope pointing. Over the 3 nights, this leads to a total of $1.1$ hours of fringe integration over approximately $5$ hours of observation (about 45 pointing). Neither the AMBER nor MIDI instruments were used simultaneously, so the only interferometric data available are from FINITO.

The seeing ranged between $0.75$'' and $2.2$'' during the observations. The fringe-tracking loop was always able to lock the fringes with average performances: locking time greater than 80\% and residual OPD generally less than $0.3$\,$\mu$m RMS. The fringes were never lost for more than a second. The main difference between good and bad files are the number of fringe jumps from which the tracking loop had to recover.

Concerning the sources, well known small and large stars were observed sequentially to test the feasibility of astrophysical data processing by measuring stellar diameters. In the following we respectively call the small and large stars calibrators and targets. Their stellar parameters are summarized in Table~\ref{tab:stars}.

\begin{table}
  \centering
  \caption{Target parameters.}
  \begin{tabular}[c]{cccccccc} \hline \hline \vspace{-0.25cm}\\
    Target      & HD     & H     & Spectral & Published \dUD{} in H\\
    Name        & Number & (mag) & Type     & (mas)   \vspace{0.03cm}\\ \hline \vspace{-0.30cm}\\
    \alphaHya{} & 81797  & -0.99 & K3II     & $9.44\pm{}?$\\
    \nVel{}     & 82668  & -0.35 & K5III    & $6.89\pm{}0.08$\\
    \deltaVir{} & 112300 & -1.0  & M3III    & $9.91\pm{}0.1$\\
    \BKVir{}    & 108849 & -0.37 & M7III    & $10.73\pm{}0.23$
    \vspace{0.03cm}\\ \hline \vspace{-0.30cm}\\ 
    \aHya{}     & 73840  &  2.0  & K3III    & $2.62\pm{}0.1$\\
    \chiVir{}   & 110014 &  2.17 & K2III    & $1.95\pm{}0.02$\\
    \iotaHya{}  & 83618  &  0.95 & K2.5III  & $3.39\pm{}0.05$\\
    \epsCrt{}   & 99167  &  1.19 & K5III    & $3.45\pm{}0.04$\\
  \end{tabular}
  \flushleft
  \footnotesize{Note: Spectral types and H-band magnitudes have been extracted from the Simbad and 2MASS database, and diameters from CHARM2 catalog \citep{Richichi-2005feb}.}
  \label{tab:stars}
\end{table}

\subsection{Data reduction}
\label{sec:DataReduction}
\paragraph{Step 1: } 

\begin{figure*} % plots.i
  \hspace{0.2cm}\includegraphics[scale=1]{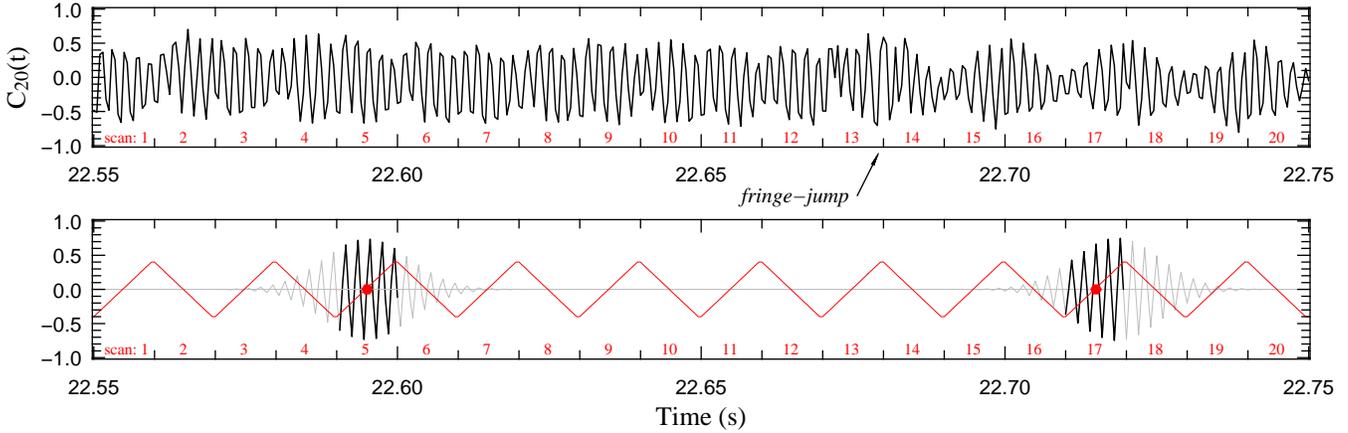}
  \centering \caption{\label{fig:fringes} Example of a FINITO real-time fringe signal $C_{20}(t)$ (upper panel) on a short sequence of $0.2$s, with its associated FMU modulation (red curve in lower panel). The FMU modulation was 5 fringes in amplitude, with 4 readouts per fringes, and $0.5$ms per readout, leading to $10$ms per scan. The scan numbering is given at the bottom of each plot. Gray fringes in the lower panel are the theoretical $H-$band fringe-packet of scans~5 and~17, assuming a infinite linear modulation ramp. The part of the packet that is effectively modulated is over-plotted in black. This pattern is replicated for each scan (mirrored in time when the FMU slope is negative). After the fringe jump ($t=22.68$s), the fringe-packet appears shifted with respect to the middle of the modulation window. This explains the apparent oscillation in the envelope of signal $C_{20}(t)$.}
\end{figure*}

The FINITO raw data are processed in real time in order to compute the correction send to the Delay Lines. In this process, the two raw interferometric outputs are corrected and calibrated from the photometric fluctuations by the following classical formula (here for beams 2-0):
\begin{eqnarray}
  \label{eq:norm}
  C_{20}^+ = 
  \frac{I_{20}^+ \;-\; \kappa_{2\to{}20^+}\,P_{2} \;-\; \kappa_{0\to{}20^+}\,P_{0}} 
  {2\sqrt{\kappa_{2\to{}20^+} \, \kappa_{0\to{}20^+} \, P_{2}\, P_{0}}} \\
  C_{20}^- = 
  \frac{I_{20}^- \;-\; \kappa_{2\to{}20^-}\,P_{2} \;-\; \kappa_{0\to{}20^-}\,P_{0}} 
  {2\sqrt{\kappa_{2\to{}20^-} \, \kappa_{0\to{}20^-} \, P_{2}\, P_{0}}}
\end{eqnarray}
The $\kappa{}$-matrix is the internal transmission ratio between the photometric and the interferometric outputs. It is calibrated each time we change star by measuring the flux at the outputs while observing one beam at a time.

The two cleaned signals are then subtracted to enhance the signal-to-noise ratio (SNR):
\begin{equation}
  \label{eq:fringe_sum}
  C_{20} =  \frac{C_{20^+} - C_{20^-}}{2}
\end{equation}
By doing so, all the correlated noises between the two outputs should disappear, while the fringes, perfectly opposed in phase, are amplified. The resulting signal $C_{20}(t)$ is processed in real time to compute the phase residual and the group-delay, which is used to feed the fringe-tracking loop. It is accessible on the RMN and can be stored into FITS files by the RMNrec machine. An example of fringe sequence is displayed in Fig.~\ref{fig:fringes}. For illustration purpose, we chose a sequence with an obvious fringe jump (at $t=22.68$s). In the last 7 scans, the fringe-packet was not centered inside the modulation window. Those fringe jumps are caused by temporary flux dropouts and can occur at unpredictable times. Between a fringe jump and the time the OPDC compensates for it, the average fringe contrast is obviously reduced. All these scans should be efficiently detected and discarded before computing the average visibility.

\paragraph{Step 2:}
We used two simple criteria for the scan selection: first the fringe-tracking loop should be closed during the complete scan, and secondly the fringe-packet maximum should be less than one fringe from the scan center. Additionally, the scans were weighted according to their average SNR. To do this, we configured the RMNrec to store all required OPDC signals (loop status, fringe-packet position, SNR).

\begin{figure} % plots.i
  \hspace{0.15cm}\includegraphics[scale=1]{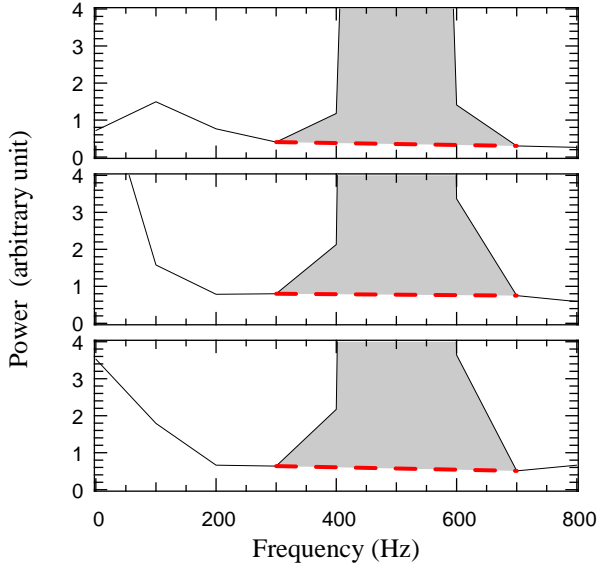}
  \centering \caption{\label{fig:psd} Average power spectral density of typical FINITO sequences of 30s, after scan selection and weighting (see Step~3). The square visibility is obtained by integrating the fringe power around 500Hz (filled region), after removing the bias (dashed red curve).}
\end{figure}

We then computed the power spectral density (PSD) of each of the accepted scans, averaged all of them, and integrated the power around the expected fringe frequency:
\begin{equation}
  \label{eq:dsp}
  \vis{}^2_{raw} = \int_{300\mathrm{Hz}}^{700\mathrm{Hz}} <\,|\widetilde{C_{a}(t)}|^2>\;\mathrm{d}f \;\;-\;\; \mathrm{bias}
\end{equation}
where $<>$ denotes the average over the accepted scans. This power is a quadratic estimation of the fringe visibility, assuming the bias is properly removed. In our data set, we estimated it by interpolating the values at the edges of the fringe peak; see Fig.~\ref{fig:psd} for typical PSDs with their bias removed. We decided to integrate the fringe power in the frequency range 300-700Hz to take into account the small amount of energy spread due to the shape of the fringe-packet in the direct space (broad-band fringes). We checked that integrating over a wider range 200-800Hz does not significantly change the final calibrated visibilities.

\begin{figure} % plots.i
  \hspace{0.15cm}\includegraphics[scale=1]{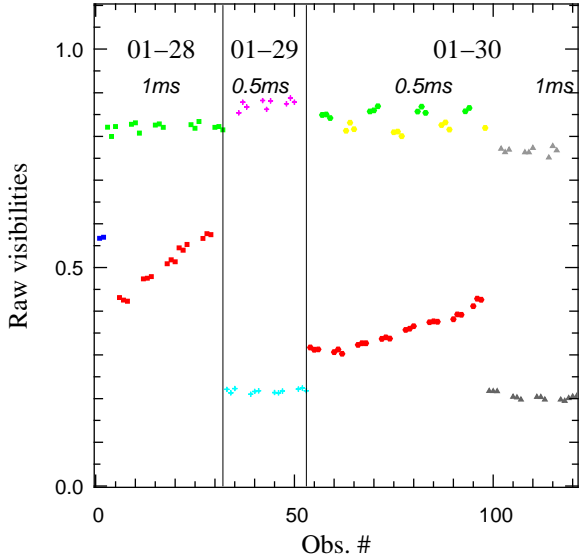}
  \centering \caption{\label{fig:data} Raw visibilities for all observations, along with the date (month-day) and the detector integration time (italic) information. See Fig.~\ref{fig:diam} for the color code of each source.}
\end{figure}

All reduced visibility points $\vis{}_{raw}$ are displayed in Fig.~\ref{fig:data}. On the largest stars, the super-synthesis effect is clearly visible, making the baseline lengths (and therefore the visibilities) slowly change over time. However, these raw visibilities cannot be used directly for estimating the stellar diameters. They have to be calibrated from the interferometric response of the whole instrumental chain, also called the transfer function.

\begin{figure} % plots.i
  \hspace{0.15cm} \includegraphics[scale=1]{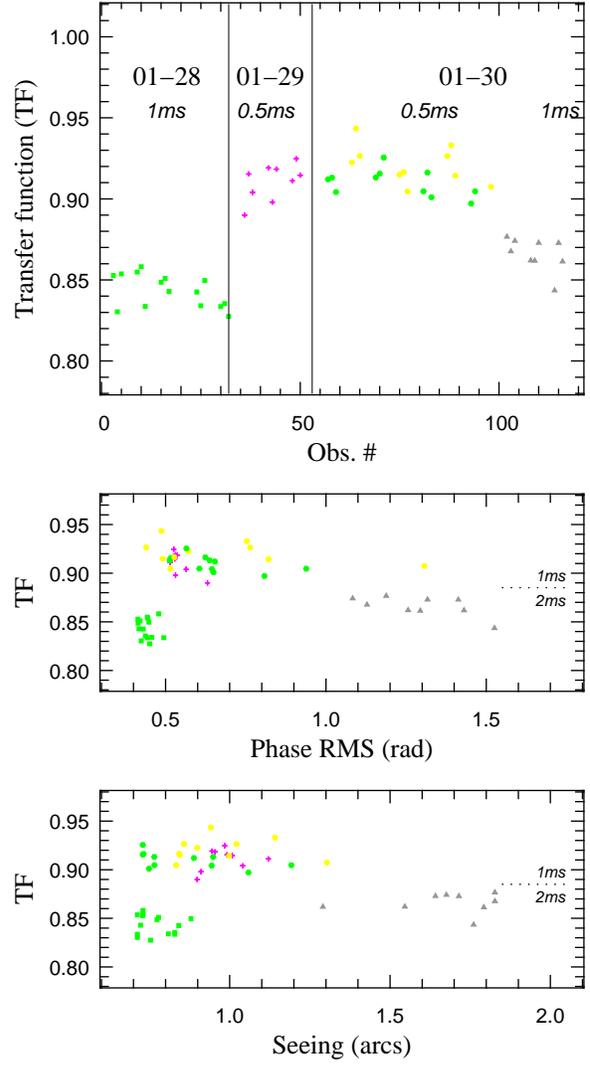}
  \centering \caption{\label{fig:tf} Top: Transfer function computed for each calibrator observation, along with the date (month-day) and the detector integration time (italic) information. Middle and bottom: same transfer function points but plotted versus the FINITO residual phase RMS (accepted scans only) and versus the Paranal DIMM seeing. See Fig.~\ref{fig:diam} for the color code of each source.}
\end{figure}

\begin{figure} % plots.i
  \hspace{0.15cm}\includegraphics[scale=1]{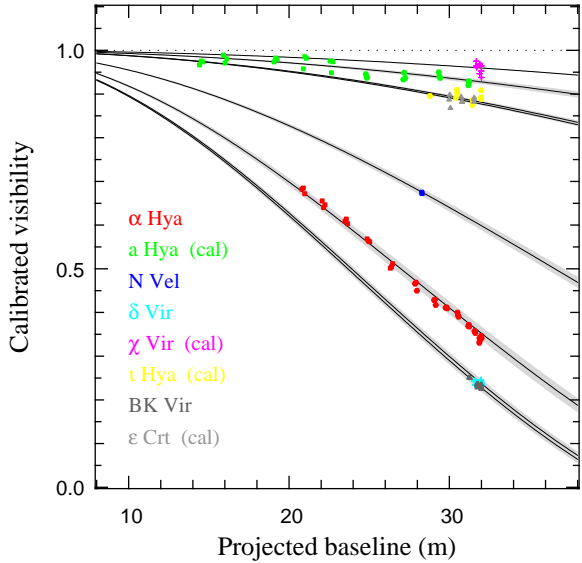}
  \centering \caption{\label{fig:diam} Calibrated visibilities of all observations plotted versus the baseline length. The solid lines are the literature diameters used to compute the transfer function (for the calibrators), or the best-fitted diameters of the data set (for the targets). The shaded regions show the diameter uncertainties. Symbols are for the different setups, defined as same night and same FINITO integration time (DIT).}
\end{figure}

\paragraph{Step 3: } We computed the instantaneous transfer function \tf{} by dividing the observed visibilities measured on each calibrators with those expected according to the diameters of Table~\ref{tab:stars}:
\begin{equation}
  \label{eq:tf}
  \tf{} =  \frac{\vis{}_{raw}}{\airy(\dUD{}\,\base/\lambda)}\;.
\end{equation}
In this well-known formula, $\base$ is the baseline length of each visibility point $\vis{}$, $\lambda{}$ the effective wavelength, and $\airy{}$ the Airy function (Fourier transform of a uniform disk). In the following, we assume an effective wavelength of $1.625\mu{}$m, as measured by Bauvir et al. (private communication). Results are plotted in Fig.~\ref{fig:tf}. Given the overall stability of the transfer function, we decided to average all points of a given setup, defined as the same night and the same detector integration time (DIT). We used this average transfer function to calibrate all observed visibilities of the corresponding setup:
\begin{equation}
  \label{eq:calib}
  \vis{} =  \frac{\vis{}_{raw}}{<\tf{}>}\,.
\end{equation}
The resulting calibrated visibilities are plotted against the baseline length in Fig.~\ref{fig:diam}. To extract uniform disk diameters \dUD{} for each target, we fit the full $\vis(\base{})$ data set with the expression $\airy(\dUD{}\,\base/\lambda)$. Measured uniform disk diameters are summarized in Table~\ref{tab:diam}, as well as comparison of previously published values.

%%%%%%%%%%%%%%%%%%%%%%%%%%%%%%%%%%%%%%%%%%%%%%%%%%%%%%%%%%%%%%%%%%%
\section{Results and discussion}
\label{sec:resultsAndDiscussion}

All four measured diameters are within $5\%$ of the previously published diameters. %The maximum difference (\nVel{}) corresponds to $3.7\sigma{}$ of the published accuracy. If real, potential discrepancies can be explained by a combination of (i) pulsations or patchy surface, since all observed targets are variable giants ; (ii) spectral dependency of the diameter; and/or (iii) limb darkening effect, since some of our observations have relatively high spatial frequencies.Most importantly, and.
This was the goal of the study: the good agreement proves the capability of the FINITO/OPDC/RMNrec facility to provide meaningful interferometric visibility measurements. Interestingly, the transfer function is stable over several hours. It is one of the most important parameters in obtaining robust visibility measurements.

\begin{table}
  \centering
  \caption{Published versus measured diameters.}
  \begin{tabular}[c]{cccccc} \hline \hline \vspace{-0.25cm}\\
    Target      & Published \dUD{} & Method & Lbd      & Measured \dUD{} & Nb. of \\
    Name        &           (mas)  & ($^1$) & ($\mu$m) & in H (mas)      & pointing 
    \vspace{0.03cm}\\ \hline \vspace{-0.30cm}\\
    \alphaHya{} & $9.09\pm{}0.09$  & LBI &  $0.8$  & $8.95 \pm{}0.15$ & $13$ \\
    \nVel{}     & $6.89\pm{}0.08$  & IND &  $1.6$  & $6.6  \pm{}?   $ & $1$  \\
    \deltaVir{} & $9.91\pm{}0.10$  & LBI &  $0.8$  & $10.1 \pm{}0.1 $ & $4$  \\
    \BKVir{}    & $10.73\pm{}0.23$ & LBI &  $2.2$  & $10.2 \pm{}0.1 $ & $5$  \\
  \end{tabular}
  \flushleft
  \label{tab:diam}
{ \footnotesize $^1$ Indirect (IND) or long baseline interferometry (LBI) estimation.}
\end{table}

\subsection{Visibility accuracy}

Total uncertainty on our visibility points is difficult to estimate. We used the bootstrapping method to compute the contribution of the statistical dispersion inside each file. This technique has the important advantage of not assuming a Gaussian distribution of the measurements. Resulting statistical errors are very small, on the order of $\pm0.1\%$, twhich is inside the symbol size of Fig.~\ref{fig:tf}. This does not explain the observed dispersion between consecutive data points. The total uncertainty is therefore dominated by systematic effects, that is to say, errors affecting all scans in the same ways, or bias in the averaging process. We see at least four potential origins:

\begin{enumerate}
\item In long baseline interferometry from the ground, the transfer function depends on the atmospheric turbulence strength. To explain the observed variations, we should invoke fast changes in the atmospheric turbulence at the level of one minute or less. But we do not see any correlations when plotting the transfer function estimations against the seeing or the fringe-jitter (see Fig.~\ref{fig:tf}), the latter being a good proxy for the coherence time. Moreover, we expect FINITO to be robust against turbulence fluctuations because it combines the use of single-mode fibers, high sampling rate, and real-time piston correction.
\item The $\kappa$-matrix and its associated errors are not stored in the FITS files produced by RMNrec, and are not propagated through Eqs.~\ref{eq:norm} to \ref{eq:calib}. Because this $\kappa$-matrix is only recomputed at each pointing, any errors on it would lead to systematic offsets \emph{between} sequence of 3 consecutive files on a given star. If present, such behaviors are not dominant in Fig.~\ref{fig:tf}, demonstrating that errors on the $\kappa$-matrix alone cannot explain the observed discrepancies.
\item As explained in Step~2 of Sect.~\ref{sec:DataReduction}, we discarded the scans where the central fringe was not centered in the modulated window. The efficiency of this selection process is correlated to the accuracy on the fringe-packet center, which may depend on the observing conditions.
\item As seen in the PSD plotted in Fig.~\ref{fig:psd}, the fringe power around 500Hz is probably contaminated by a low-frequency signal. Its exact spectral shape is unknown and looks unstable, making it difficult to fit correctly. In the data reduction, we used a linear interpolation between arbitrary frequencies to estimate and remove it (see Step~2 of Sect.~\ref{sec:DataReduction}). This may have introduced an additional error in the integrated fringe power.
\end{enumerate}

\subsection{Perspectives toward higher accuracy}
\label{sec:perspectivesPrecise}

At the time of our observations, RMNrec was only able to store the signal $C_{20}(t)$, not the FINITO raw signal themselves. Therefore, as explained in Sect.~\ref{sec:DataReduction}, the results presented in this paper are based on data that were previously processed in real time by Eqs.~\ref{eq:norm} and~\ref{eq:fringe_sum}. Data reduction would highly benefit from a logging of the raw FINITO data. For instance, photometric calibration during post-processing can make use of Wiener Filtering, which is not possible in real time since this filter is not causal. Additionally, logging the $\kappa$-matrix and associated errors is mandatory to check its stability and its effective contribution into the final error.

FINITO is a fringe tracker, meaning that fringes are locked in phase. Using coherent average over several scans (instead of power-average as in this paper) can reduce the impact of the bias on the final result. On the other hand, the transfer function of coherent-average depends on the fringe-tracking loop performances. All together, this approach is worth looking at since it may be a useful trade-off for fainter stars, where bias subtraction becomes more complicated. We plan to do so as soon as the logging of the OPDC/FINITO data becomes a standard procedure.

\subsection{Perspectives toward fainter targets}
\label{sec:perspectivesFaint}

More generally, we should discuss the expected performances on fainter targets. Our main concern is related to Step~2 of the proposed data-reduction (discarding the fringe packets that were not properly centered inside the modulated window). Indeed, the efficiency of such scan selection, while good on high-SNR fringes, most probably degrades rapidly on noisy data. The simplest and most robust solution is to modulate an OPD-range significantly wider than the fringe-packet size, as with the instrument VINCI for instance \citep{Kervella-2004oct,LeBouquin-2004sep}. However, the FINITO fringe-tracking capability would be lost because the phase measurements become extremely noisy while outside the fringe-packet. Performing packet-tracking would still be possible, but this prohibits the use of long integration on the scientific camera (required for medium and high spectral resolution modes of AMBER). Because of its potential scientific interest on faint targets, we are looking forward to implementing long scans and group-tracking as an additional mode of FINITO.

%%%%%%%%%%%%%%%%%%%%%%%%%%%%%%%%%%%%%%%%%%%%%%%%%%%%%%%%%%%%%%%%%%%
\section{Conclusions}

We have demonstrated the possibility of post-processing the real-time fringe-tracking data of the Very Large Telescope Interferometer for astronomical purposes. We proposed a simple strategy for reducing the data, based on scan selection and power integration. We were able to recover the $H$-band diameters of four already-known giant stars. The good general agreement proved the capability of the FINITO/OPDC/RMNrec facility to provide meaningful interferometric visibility measurements.

Interestingly, the transfer function level is dominated by the integration time. It is stable over several hours, with possible significant changes from one day to the next. We have not been able to find any obvious correlation with the atmospheric seeing and fringe-tracking performances, even if we agree that our dataset does not allow us to draw definite conclusions. Post-processing the FINITO/OPDC data therefore provides a rough, but robust, estimation of the $H$-band absolute visibility even without the observation of a calibration star close in time. This complements the scientific observations very well, generally done in another spectral range (near-IR for AMBER, and $N$-band for MIDI) and with higher spectral resolution (up to $10\,000$ for AMBER), but with only poor absolute calibration of the continuum visibility. We emphasize that this additional, crucial measure is obtained simultaneously and without any additional overhead time.

This demonstration is also promising in the context of precision interferometry at the VLTI. We were able to reach accuracy of $\pm{}2\%$ on the raw visibilities with a simple data reduction method. We showed that the overall accuracy is most probably limited by systematic errors. Statistical errors only about $\pm0.1\%$ for 30s integration on $\mathrm{H}=2^{\rm m}$ unresolved targets with the 1.8m Auxiliary Telescopes.

Based on the success of this demonstration, we now plan to routinely log and archive the VLTI fringe-tracking data. In this context, we emphasize the need for a dedicated public data reduction pipeline.

\begin{acknowledgements} 
  The authors want to warmly thank the VLTI team. This work is based on technical observations made with the ESO telescopes. It made use of the Smithsonian/NASA Astrophysics Data System (ADS) and of the Centre de Donnees astronomiques de Strasbourg (CDS). All calculations and graphics were performed with the freeware \texttt{Yorick}\footnote{\texttt{http://yorick.sourceforge.net}}. JB Le~Bouquin wants to thank JP. Berger and S. Miner for their revisions of the manuscript.
\end{acknowledgements}

%%%%%%%%%%%%%%%%%%%%%%%%%%%%%%%%%%%%%%%%%%%%%%%%%%%%%%%%%%%%%%%%%%%

% \bibliographystyle{/Users/jlebouqu/Tex/AandA/aa}  % style for A&A
% \bibliography{/Users/jlebouqu/Biblio/BibTex/all}   % BiTex files

\begin{thebibliography}{13}
\expandafter\ifx\csname natexlab\endcsname\relax\def\natexlab#1{#1}\fi

\bibitem[{{Domiciano de Souza} {et~al.}(2007){Domiciano de Souza}, {Driebe},
  {Chesneau}, {Hofmann}, {Kraus}, {Miroshnichenko}, {Ohnaka}, {Petrov},
  {Preisbisch}, {Stee}, {Weigelt}, {Lisi}, {Malbet}, \&
  {Richichi}}]{Domiciano-2007mar}
{Domiciano de Souza}, A., {Driebe}, T., {Chesneau}, O., {et~al.} 2007, \aap,
  464, 81

\bibitem[{{Gai} {et~al.}(2004){Gai}, {Menardi}, {Cesare}, {Bauvir}, {Bonino},
  {Corcione}, {Dimmler}, {Massone}, {Reynaud}, \& {Wallander}}]{Gai-2004oct}
{Gai}, M., {Menardi}, S., {Cesare}, S., {et~al.} 2004, in New Frontiers in
  Stellar Interferometry, Proc. of SPIE, Vol. 5491. Edited by Wesley A. Traub.
  Bellingham, WA: The International Society for Optical Engineering, 2004.,
  p.528, ed. W.~A. {Traub}, Vol. 5491, 528--+

\bibitem[{{Kervella} {et~al.}(2004){Kervella}, {S{\'e}gransan}, \& {Coud{\'e}
  du Foresto}}]{Kervella-2004oct}
{Kervella}, P., {S{\'e}gransan}, D., \& {Coud{\'e} du Foresto}, V. 2004, \aap,
  425, 1161

\bibitem[{{Le Bouquin} {et~al.}(2008{\natexlab{a}}){Le Bouquin}, {Abuter},
  {Bauvir}, {Bonnet}, {Haguenauer}, {di Lieto}, {Menardi}, {Morel},
  {Rantakyr{\"o}}, {Schoeller}, {Wallander}, \&
  {Wehner}}]{LeBouquin-2008spie_b}
{Le Bouquin}, J.-B., {Abuter}, R., {Bauvir}, B., {et~al.} 2008{\natexlab{a}},
  in Presented at the Society of Photo-Optical Instrumentation Engineers (SPIE)
  Conference, Vol. 7013, Society of Photo-Optical Instrumentation Engineers
  (SPIE) Conference Series

\bibitem[{{Le Bouquin} {et~al.}(2008{\natexlab{b}}){Le Bouquin}, {Bauvir},
  {Haguenauer}, {Sch{\"o}ller}, {Rantakyr{\"o}}, \&
  {Menardi}}]{LeBouquin-2008apr}
{Le Bouquin}, J.-B., {Bauvir}, B., {Haguenauer}, P., {et~al.}
  2008{\natexlab{b}}, \aap, 481, 553

\bibitem[{{Le Bouquin} {et~al.}(2004){Le Bouquin}, {Rousselet-Perraut}, {Kern},
  {Malbet}, {Haguenauer}, {Kervella}, {Schanen}, {Berger}, {Delboulb{\'e}},
  {Arezki}, \& {Sch{\"o}ller}}]{LeBouquin-2004sep}
{Le Bouquin}, J.-B., {Rousselet-Perraut}, K., {Kern}, P., {et~al.} 2004, \aap,
  424, 719

\bibitem[{{Meilland} {et~al.}(2007){Meilland}, {Stee}, {Vannier}, {Millour},
  {Domiciano de Souza}, {Malbet}, {Martayan}, {Paresce}, {Petrov}, {Richichi},
  \& {Spang}}]{Meilland-2007mar}
{Meilland}, A., {Stee}, P., {Vannier}, M., {et~al.} 2007, \aap, 464, 59

\bibitem[{{Millour} {et~al.}(2007){Millour}, {Petrov}, {Malbet}, {Tatulli},
  {Duvert}, {Zins}, {Altariba}, {Vannier}, {Hernandez}, \&
  {Causi}}]{Millour-2007may}
{Millour}, F., {Petrov}, R., {Malbet}, F., {et~al.} 2007, ArXiv e-prints, 705

\bibitem[{{Petrov} {et~al.}(2007){Petrov}, {Malbet}, {Weigelt}, {Antonelli},
  {Beckmann}, {Bresson}, {Chelli}, {Dugu{\'e}}, {Duvert}, {Gennari},
  {Gl{\"u}ck}, {Kern}, {Lagarde}, {Le Coarer}, {Lisi}, {Millour}, {Perraut},
  {Puget}, {Rantakyr{\"o}}, {Robbe-Dubois}, {Roussel}, {Salinari}, {Tatulli},
  {Zins}, {Accardo}, {Acke}, {Agabi}, {Altariba}, {Arezki}, {Aristidi},
  {Baffa}, {Behrend}, {Bl{\"o}cker}, {Bonhomme}, {Busoni}, {Cassaing},
  {Clausse}, {Colin}, {Connot}, {Delboulb{\'e}}, {Domiciano de Souza},
  {Driebe}, {Feautrier}, {Ferruzzi}, {Forveille}, {Fossat}, {Foy},
  {Fraix-Burnet}, {Gallardo}, {Giani}, {Gil}, {Glentzlin}, {Heiden},
  {Heininger}, {Hernandez Utrera}, {Hofmann}, {Kamm}, {Kiekebusch}, {Kraus},
  {Le Contel}, {Le Contel}, {Lesourd}, {Lopez}, {Lopez}, {Magnard}, {Marconi},
  {Mars}, {Martinot-Lagarde}, {Mathias}, {M{\`e}ge}, {Monin}, {Mouillet},
  {Mourard}, {Nussbaum}, {Ohnaka}, {Pacheco}, {Perrier}, {Rabbia}, {Rebattu},
  {Reynaud}, {Richichi}, {Robini}, {Sacchettini}, {Schertl}, {Sch{\"o}ller},
  {Solscheid}, {Spang}, {Stee}, {Stefanini}, {Tallon}, {Tallon-Bosc}, {Tasso},
  {Testi}, {Vakili}, {von der L{\"u}he}, {Valtier}, {Vannier}, \&
  {Ventura}}]{Petrov-2007mar}
{Petrov}, R.~G., {Malbet}, F., {Weigelt}, G., {et~al.} 2007, \aap, 464, 1

\bibitem[{{Richichi} {et~al.}(2005){Richichi}, {Percheron}, \&
  {Khristoforova}}]{Richichi-2005feb}
{Richichi}, A., {Percheron}, I., \& {Khristoforova}, M. 2005, \aap, 431, 773

\bibitem[{{Sch{\"o}ller} {et~al.}(2006){Sch{\"o}ller}, {Argomedo}, {Bauvir},
  {Blanco-Lopez}, {Bonnet}, {Brillant}, {Cantzler}, {Carstens}, {Caruso},
  {Choque-Cortez}, {Derie}, {Delplancke}, {Di Lieto}, {Dimmler}, {Durand},
  {Ferrari}, {Galliano}, {Gitton}, {Gilli}, {Glindemann}, {Guniat}, {Guisard},
  {Haddad}, {Haguenauer}, {Housen}, {Hudepohl}, {Hummel}, {Kaufer},
  {Kiekebusch}, {Koehler}, {Le Bouquin}, {Leveque}, {Lidman}, {Mardones},
  {Menardi}, {Morel}, {Mornhinweg}, {Nicoud}, {Percheron}, {Petr-Gotzens},
  {Duc}, {Puech}, {Ramirez}, {Rantakyr{\"o}}, {Richichi}, {Rivinius},
  {Sandrock}, {Somboli}, {Spyromilio}, {Stefl}, {Suc}, {Tamai}, {Tapia},
  {Vannier}, {Vasisht}, {Wallander}, {Wehner}, {Wittkowski}, \&
  {Zagal}}]{Scholler-2006jul}
{Sch{\"o}ller}, M., {Argomedo}, J., {Bauvir}, B., {et~al.} 2006, in Presented
  at the Society of Photo-Optical Instrumentation Engineers (SPIE) Conference,
  Vol. 6268, Advances in Stellar Interferometry. Edited by Monnier, John D.;
  Sch{\"o}ller, Markus; Danchi, William C.. Proc. of the SPIE, Vol. 6268, pp.
  62680L (2006).

\bibitem[{{Weigelt} {et~al.}(2007){Weigelt}, {Kraus}, {Driebe}, {Petrov},
  {Hofmann}, {Millour}, {Chesneau}, {Schertl}, {Malbet}, {Hillier}, {Gull},
  {Davidson}, {Domiciano de Souza}, {Antonelli}, {Beckmann}, {Bresson},
  {Chelli}, {Dugu{\'e}}, {Duvert}, {Gennari}, {Gl{\"u}ck}, {Kern}, {Lagarde},
  {Le Coarer}, {Lisi}, {Perraut}, {Puget}, {Rantakyr{\"o}}, {Robbe-Dubois},
  {Roussel}, {Tatulli}, {Zins}, {Accardo}, {Acke}, {Agabi}, {Altariba},
  {Arezki}, {Aristidi}, {Baffa}, {Behrend}, {Bl{\"o}cker}, {Bonhomme},
  {Busoni}, {Cassaing}, {Clausse}, {Colin}, {Connot}, {Delboulb{\'e}},
  {Feautrier}, {Ferruzzi}, {Forveille}, {Fossat}, {Foy}, {Fraix-Burnet},
  {Gallardo}, {Giani}, {Gil}, {Glentzlin}, {Heiden}, {Heininger}, {Hernandez
  Utrera}, {Kamm}, {Kiekebusch}, {Le Contel}, {Le Contel}, {Lesourd}, {Lopez},
  {Lopez}, {Magnard}, {Marconi}, {Mars}, {Martinot-Lagarde}, {Mathias},
  {M{\`e}ge}, {Monin}, {Mouillet}, {Mourard}, {Nussbaum}, {Ohnaka}, {Pacheco},
  {Perrier}, {Rabbia}, {Rebattu}, {Reynaud}, {Richichi}, {Robini},
  {Sacchettini}, {Sch{\"o}ller}, {Solscheid}, {Spang}, {Stee}, {Stefanini},
  {Tallon}, {Tallon-Bosc}, {Tasso}, {Testi}, {Vakili}, {von der L{\"u}he},
  {Valtier}, {Vannier}, {Ventura}, {Weis}, \& {Wittkowski}}]{Weigelt-2007mar}
{Weigelt}, G., {Kraus}, S., {Driebe}, T., {et~al.} 2007, \aap, 464, 87

\bibitem[{{Wittkowski} {et~al.}(2008){Wittkowski}, {Boboltz}, {Driebe}, {Le
  Bouquin}, {Millour}, {Ohnaka}, \& {Scholz}}]{Wittkowski-2008feb}
{Wittkowski}, M., {Boboltz}, D.~A., {Driebe}, T., {et~al.} 2008, \aap, 479, L21

\end{thebibliography}
 
%%%%%%%%%%%%%%%%%%%%%%%%%%%%%%%%%%%%%%%%%%%%%%%%%%%%%%%%%%%%%%%%%%%
% \newpage
% \appendix

%%%%%%%%%%%%%%%%%%%%%%%%%%%%%%%%%%%%%%%%%%%%%%%%%%%%%%%%%%%%%%%%%%%
%% END DOCUMENT 
%%%%%%%%%%%%%%%%%%%%%%%%%%%%%%%%%%%%%%%%%%%%%%%%%%%%%%%%%%%%%%%%%%%%%%%%%%%
\end{document}